\documentclass[sigconf]{acmart}

\setcopyright{none}
\renewcommand\footnotetextcopyrightpermission[1]{}

\usepackage{booktabs}
\usepackage{array}
\usepackage{pifont}
\newcommand{\yes}{\ding{51}}
\newcommand{\no}{\ding{55}}
\usepackage{tikz}
\usetikzlibrary{arrows.meta,positioning,calc}

\makeatletter
\def\@ACM@flataffil#1#2{%
  \begingroup
    \setkeys{@ACM@affiliation@}{obeypunctuation=false}%
    \setkeys{@ACM@affiliation@}{#1}\@ACM@resetaffil
    #2\@ACM@checkaffil
  \endgroup}
\def\@mkauthors@flat{%
  \def\@currentauthors{}%
  \def\@currentaffiliations{}%
  \def\@currentemails{}%
  \global\let\and\relax
  \def\@author##1{%
    \ifx\@currentauthors\@empty
      \gdef\@currentauthors{##1}%
    \else
      \g@addto@macro{\@currentauthors}{\and ##1}%
    \fi}%
  \def\email##1##2{%
    \ifx\@currentemails\@empty
      \gdef\@currentemails{\nolinkurl{##2}}%
    \else
      \g@addto@macro{\@currentemails}{\and\nolinkurl{##2}}%
    \fi}%
  \def\affiliation##1##2{%
    \def\@tempa{##2}\ifx\@tempa\@empty\else
      \ifcsname @ACM@seenaffil@\detokenize{##1|##2}\endcsname\else
        \expandafter\gdef\csname @ACM@seenaffil@\detokenize{##1|##2}\endcsname{}%
        \ifx\@currentaffiliations\@empty
          \gdef\@currentaffiliations{\@ACM@flataffil{##1}{##2}}%
        \else
          \g@addto@macro{\@currentaffiliations}{%
            \smallskip\@ACM@flataffil{##1}{##2}}%
        \fi
      \fi
    \fi}%
  \global\setbox\mktitle@bx=\vbox{%
    \noindent\unvbox\mktitle@bx\par\medskip
    \addresses
    \begingroup
      \centering
      \@authorfont\andify\@currentauthors\@currentauthors\par
      \medskip
      \@affiliationfont\@currentaffiliations\par
      \ifx\@currentemails\@empty\else
        \smallskip\andify\@currentemails\@currentemails\par
      \fi
    \endgroup
    \medskip}%
}
\renewcommand\@mkauthors{\begingroup\hsize=\textwidth\@mkauthors@flat\endgroup}
\makeatother

\tikzset{
  pstage/.style={draw=black!45, rounded corners=2pt, fill=black!4,
                 align=center, font=\scriptsize,
                 inner xsep=4pt, inner ysep=3pt,
                 text width=0.74\columnwidth},
  wstage/.style={draw=black!45, rounded corners=2pt, fill=black!4,
                 align=center, font=\scriptsize,
                 inner xsep=4pt, inner ysep=3.5pt, text width=1.55cm},
  rstage/.style={draw=black!30, rounded corners=2pt, fill=black!2,
                 align=center, font=\scriptsize\itshape,
                 inner xsep=4pt, inner ysep=3.5pt, text width=1.9cm},
  sstage/.style={draw=black!45, rounded corners=2pt, fill=black!4,
                 align=center, font=\scriptsize,
                 inner xsep=4pt, inner ysep=3.5pt, text width=2.05cm},
  band/.style={draw=black!35, dashed, rounded corners=2pt, fill=black!2,
               align=center, font=\scriptsize\itshape, inner sep=4pt},
  parrow/.style={-{Latex[length=4pt,width=3.5pt]}, semithick, black!55},
  farrow/.style={-{Latex[length=4pt,width=3.5pt]}, semithick, black!40, dashed}
}
\newcommand{\stepno}[1]{\textbf{[#1]}\;}

\newcommand{\arw}{\ensuremath{\,\rightarrow\,}\allowbreak}
\begin{document}

\title{From PyTorch to the NPU: LLM-Agent-Driven Model Conversion
       Across Heterogeneous Inference Runtimes}
\subtitle{AI for Porting and Inferencing on OpenVINO, RKNN, TensorRT, and ONNX}

%
\newcommand{\affQC}{\texorpdfstring{\textsuperscript{1}}{}}
\newcommand{\affNCKU}{\texorpdfstring{\textsuperscript{2}}{}}

\author{Jianhao Su\affQC}
\author{Zhanwei Wu\affQC}
\affiliation{%
  \institution{\affQC Qualcomm Technologies, Inc.}
  \country{}}

\author{Chia-Heng Tu\affNCKU}
\affiliation[obeypunctuation=true]{%
  \institution{\affNCKU National Cheng Kung University},
  \city{Tainan},
  \country{Taiwan}}

\author{ShengTing Huang\affQC}
\affiliation{%
  \institution{\affQC Qualcomm Technologies, Inc.}
  \country{}}

\renewcommand{\shortauthors}{J. Su et al.}

\begin{abstract}
Edge AI model deployment is a multi-stage engineering process involving model
conversion, operator compatibility handling, runtime integration, and precision
verification. While prior work has demonstrated agent-based automation for
Qualcomm AI Runtime, the broader edge inference runtime ecosystem, including
Intel OpenVINO, Rockchip RKNN, NVIDIA TensorRT, and ONNX Runtime, presents
distinct toolchains and optimization strategies.

This paper extends AIPC (AI Porting Conversion, an LLM agent-driven methodology
for AI model deployment automation previously demonstrated on Qualcomm AI
Runtime) to multi-runtime scenarios, proposing an LLM agent-driven approach for
automated single-model-to-single-runtime deployment across heterogeneous
inference backends, such as Intel OpenVINO, Rockchip RKNN, NVIDIA TensorRT, and
ONNX Runtime. We decompose the edge AI deployment into standardized, verifiable
stages, and inject runtime-specific domain knowledge into the agent execution
flow through agent skills, auxiliary scripts, and staged verification loops.

Using representative vision models, we demonstrate that agent-based deployment
can complete the conversion from a PyTorch model to its executable inference,
targeting OpenVINO for x86/NPU, RKNN for RK3588, TensorRT for NVIDIA GPU, and
ONNX Runtime for Qualcomm NPU with a focus on FP16 precision deployment
feasibility verification. The contributions of this paper primarily lie in
providing multi-runtime deployment engineering practice experience, toolchain
mapping analysis, a layout-adaptation and inference-replacement layer that
removes manual transpose insertion from the agent's repair burden, and an
empirical characterization of agent deviation behavior under structured
knowledge injection, rather than large-scale systematic benchmarking or
cross-runtime operator repair strategy comparison.
\end{abstract}


\maketitle

\thispagestyle{empty}
\pagestyle{plain}
\fancyhf{}
\fancyfoot[C]{\thepage}
\renewcommand{\headrulewidth}{0pt}

\section{Introduction}
As edge AI applications expand from image classification to object detection and
multimodal understanding, model deployment has become a critical engineering
stage determining whether a system can take advantage of underlying hardware.
The deployment environment is further complicated by the diversity of target
hardware and its runtime software, and each impose different constraints on
model structure, operator support, quantization schemes, and runtime APIs.
Significant gaps remain between models and their associated code on these
heterogeneous runtimes.

In practice, deployment goes far beyond model export. An example of a typical
model deployment flow requires ONNX conversion, operator substitution, dynamic
shape resolution, computational graph repair, runtime-specific compilation,
preprocessing and postprocessing adaptation, and precision verification. For
each runtime software, developers must understand not only the model itself but
also toolchain behavior, operator support boundaries, layout constraints, and
platform-specific differences.

Large Language Models (LLMs) have demonstrated powerful capabilities in code
generation, automatic repair, task orchestration, and tool use. Building on the
AIPC framework initially demonstrated for Qualcomm AI Runtime
(QAIRT)~\cite{aipc}, this paper investigates the following question: \emph{given
multiple deployment toolchains and verification workflows, can LLM agents
effectively take over the repetitive engineering work in
cross-heterogeneous-runtime AI model deployment, and execute bounded automatic
repair when conversion failures or runtime exceptions occur?}

The key question is not whether ``LLMs understand every hardware detail,'' but
rather: when domain knowledge is explicitly provided through agent skills,
auxiliary functions, and verification rules, can the LLM serve as a task
orchestrator and executor, transforming complex multi-runtime deployment
workflows into repeatable, verifiable, and continuously improvable automated
workflows?

The major contributions of this paper are summarized as follows.

\begin{enumerate}
\item \textbf{Multi-Runtime Deployment Design}. We generalize the AIPC agent
      methodology from a single runtime (QAIRT) to four heterogeneous inference
      backends, including Intel OpenVINO, Rockchip RKNN, NVIDIA TensorRT, and
      ONNX Runtime QNN (Qualcomm Neural Network), demonstrating the preliminary
      feasibility of the ``skill + verification loop'' design pattern across
      different toolchains.

\item \textbf{Cross-Runtime Deployment Pipeline Comparison}. We provide
      deployment pipeline mapping analysis for five runtimes (including QAIRT),
      comparing conversion modes, output formats, and toolchain characteristics
      across runtimes. This analysis lays the foundation for subsequent
      systematic comparative studies in the future.

\item \textbf{Inference Replacer for Layout Adaptation}. A drop-in replacement for ONNX Runtime is presented, called ONNXWrapper, to automatically
      load the corresponding hardware runtime model for precision comparison
      and to perform NCHW\,$\leftrightarrow$\,NHWC layout adaptation at the ONNX
      level. This removes the need for per-model manual transpose insertion and eliminates a class of layout-induced conversion failures.

\item \textbf{Empirical Characterization of Agent Deviation}. Empirical experiments have been conducted to evaluate the deviation behaviors of LLM agents with and without structured domain-knowledge injection. Through a controlled comparison of deployments, we find that agents deviate from prescribed workflows under both conditions. A careful design of verification mechanisms is needed to address this issue. 
\end{enumerate}

The remainder of this paper is organized as follows. Section~\ref{sec:bg}
describes the inference runtimes addressed in this work, related work on
agent-driven deployment, and the cross-runtime challenges that motivate our
approach. Section~\ref{sec:method} presents the agentic deployment methodology
and the ``skill + verification loop'' design pattern. Section~\ref{sec:pipe}
details the deployment pipeline, agent skill specifics, and repair actions for
each target runtime. Section~\ref{sec:case} reports two groups of case studies
on representative vision models. Section~\ref{sec:concl} concludes this work.

\section{Background}
\label{sec:bg}
This section presents the concepts relevant to this work.
Section~\ref{sec:bg:overview} introduces the common edge AI deployment pipeline.
Sections~\ref{sec:bg:ov} through~\ref{sec:bg:qairt} describe the toolchain
characteristics and deployment features of different target runtimes.
Section~\ref{sec:bg:rw} covers related work on agent-driven deployment, and
Section~\ref{sec:bg:chal} discusses the cross-runtime challenges that motivate
this study.

\subsection{Hardware Inference Runtime Overview}
\label{sec:bg:overview}
Edge AI model deployment ultimately requires execution on specific hardware
platforms, where different hardware architectures (CPU, GPU, and NPU) provide their
inference runtimes optimized for their hardware characteristics. Especially,
native hardware runtimes can fully leverage hardware parallel computing
capabilities, memory architectures, and dedicated acceleration units. It
achieves significant performance improvements and power consumption reductions.

However, these native hardware runtimes differ significantly in toolchains,
operator support, quantization schemes, and software interfaces (APIs). In terms
of precision support, different devices and runtimes have varying support for
low-precision formats. Some devices support floating-point low-precision formats
like FP16 and BF16, some only support integer quantization formats (e.g., INT8,
INT4), and others support multiple precisions simultaneously. These
hardware-level constraints directly impact deployment workflows, with Neural
Processing Unit (NPU) devices being particularly restrictive, offering far less
flexibility than CPUs and GPUs.

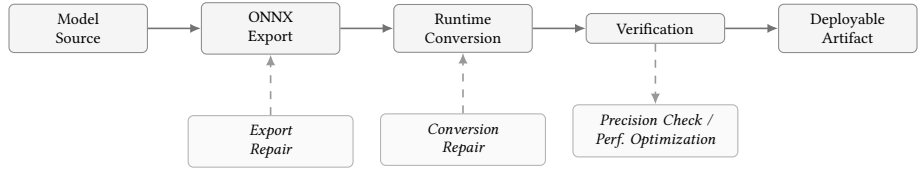
\begin{figure*}[t]
\centering
\begin{tikzpicture}[node distance=5mm and 7mm]
  \node[wstage]                    (src)  {Model\\Source};
  \node[wstage, right=of src]      (onnx) {ONNX\\Export};
  \node[wstage, right=of onnx]     (conv) {Runtime\\Conversion};
  \node[wstage, right=of conv]     (ver)  {Verification};
  \node[wstage, right=of ver]      (art)  {Deployable\\Artifact};

  \draw[parrow] (src)  -- (onnx);
  \draw[parrow] (onnx) -- (conv);
  \draw[parrow] (conv) -- (ver);
  \draw[parrow] (ver)  -- (art);

  \node[rstage, below=8mm of onnx] (er) {Export\\Repair};
  \node[rstage, below=8mm of conv] (cr) {Conversion\\Repair};
  \node[rstage, below=8mm of ver]  (pc) {Precision Check /\\Perf.\ Optimization};

  \draw[farrow] (er) -- (onnx);
  \draw[farrow] (cr) -- (conv);
  \draw[farrow] (ver) -- (pc);
\end{tikzpicture}
\caption{A typical edge AI model deployment pipeline. Solid arrows show the
forward flow; dashed arrows show the repair and verification paths that the
agent exercises when a stage fails.}
\label{fig:common}
\end{figure*}

A typical edge AI model deployment processing flow (pipeline) is illustrated in
Fig.~\ref{fig:common}. The main steps of this processing pipeline are introduced
as follows. It should confirm that the original model runs correctly and
generates reference outputs. It exports the model to ONNX~\cite{onnx}
intermediate representation as a common starting point for all runtimes, most
commonly from a PyTorch~\cite{pytorch} source model. It uses target runtime
conversion tools to generate deployable formats (quantization may be completed
as a built-in step of conversion). It verifies that the converted model's output
shapes and precision meet expectations. After deployment stabilizes (the
converted model is available), further optimization, quantization, and
performance analysis can be performed.

\subsection{Intel OpenVINO}
\label{sec:bg:ov}
Intel OpenVINO (Open Visual Inference and Neural Network
Optimization)~\cite{openvino} is Intel's toolkit specifically designed for
optimizing and deploying deep learning models on Intel hardware, supporting
CPUs, integrated GPUs, and Neural Processing Units (NPUs). Its deployment
characteristics are as follows.

\begin{itemize}
\item Operator support varies across CPU, GPU, and NPU plugins. The NPU has the
      most limited operator set.
\item Dynamic shape support varies by devices. CPUs and GPUs natively support
      dynamic shapes, while NPUs have limited dynamic shape support requiring
      workarounds, such as padding under certain conditions. Official
      pre-compiled models could be provided as alternatives to overcome this
      limitation.
\item NPU memory constraints may require model partitioning or subgraph
      offloading.
\item Model format is IR (Intermediate Representation, \texttt{.xml} +
      \texttt{.bin}). Device-related model caching mechanisms are also
      available. For example, caching compiled models to disk could be used to
      reduce initialization time on subsequent loads.
\end{itemize}

\subsection{Rockchip RKNN}
\label{sec:bg:rknn}
Rockchip RKNN (Rockchip Neural Network)~\cite{rknpu2} is the inference SDK for
Rockchip NPUs. Its deployment characteristics are as follows.

\begin{itemize}
\item Precision and operator support are strongly correlated with specific chip
      models; different RKNN chips (e.g., RV1106 vs.\ RK3566) have different
      operator sets and quantization capabilities, requiring closer attention to
      underlying hardware specifications during deployment.
\item Operator support is relatively limited, requiring custom implementations
      or CPU fallback.
\item Strict constraints on tensor shapes, layout (preferring NHWC), and
      quantization parameters.
\end{itemize}

\subsection{NVIDIA TensorRT}
\label{sec:bg:trt}
NVIDIA TensorRT~\cite{tensorrt} is a high-performance deep learning inference
optimizer and runtime for NVIDIA GPUs, supporting data center and edge
platforms. TensorRT provides two main model deployment paths.
\textbf{(1) ONNX\arw{}TensorRT} exports the model to ONNX format
first, then optimizing and compiling via the TensorRT Builder.
\textbf{(2) Torch-TensorRT} directly compiles PyTorch models into
TensorRT-optimized subgraphs via the PyTorch integration package, without manual
ONNX export, suitable for mixed deployment scenarios requiring partial PyTorch
native execution. Its deployment characteristics are as follows.

\begin{itemize}
\item Serialized engines cannot be ported across chips or platforms.
      Recompilation for the target GPU architecture is required.
\item Some ONNX operators are unsupported, requiring plugin development or
      computational graph rewriting.
\item Dynamic shape support has range limitations. TensorRT requires specifying
      minimum, optimal, and maximum values for each dynamic dimension in an
      Optimization Profile. Input shapes outside this range cannot be executed.
\item Build-time optimization (strategy selection) may be time-consuming for
      complex models.
\end{itemize}

\subsection{ONNX Runtime QNN}
\label{sec:bg:ortqnn}
ONNX Runtime~\cite{onnxruntime} is an ONNX inference engine supported by
Microsoft. This runtime supports multiple hardware backends through the
Execution Provider (EP) architecture. The QNN Execution Provider (QNN EP)
integrates ONNX Runtime with QAIRT, enabling ONNX models to run on Qualcomm
Hexagon NPUs across smartphones, notebooks, IoT, and other platforms. Its
deployment characteristics are as follows.

\begin{itemize}
\item ONNX Runtime QNN EP operator coverage depends on QNN backend support.
      Unsupported operators fall back to CPU execution, potentially causing
      significant performance degradation.
\item Compatibility between QAIRT SDK versions and ONNX Runtime versions
      requires careful management.
\item QNN EP models require fixed input dimensions.
\end{itemize}

\paragraph{Olive model optimization tool}
Microsoft Olive is a model optimization toolchain for ONNX Runtime, designed
around the concept of \textbf{Recipes} (optimization passes). That is, users
compose a series of optimization passes in declarative JSON configuration files
as recipes for FP16 conversion, static quantization (INT8/INT16 calibration),
graph-level operator fusion (\texttt{OnnxModelOptimizer},
\texttt{OrtTransformersOptimization}), QNN Context Binary generation, and
runtime tuning (\texttt{OrtPerfTuning}), with full reproducibility and version
traceability.

Instead of adopting the recipe pipeline mechanism, this work judiciously calls
Olive's \texttt{QNNPreprocess} pass, which is a preprocessing step designed for
QNN EP compatibility. Especially, it automatically replaces or decomposes
unsupported operators into equivalent supportable subgraphs, reducing the
likelihood of operator fallback to CPU when the model executes on the HTP
(Hexagon Tensor Processor) backend. This feature is a capability that
\texttt{onnxruntime.quantization} alone cannot provide. Another consideration for
adopting Olive is \textbf{future optimization flexibility}: when quantization
calibration, graph optimization, or performance tuning needs arise later, passes
can be directly added to the existing Olive calls without changing the toolchain
or restructuring deployment scripts.


\subsection{Qualcomm AI Runtime}
\label{sec:bg:qairt}
QAIRT~\cite{qnnsdk} is Qualcomm's AI inference SDK, covering inference scenarios
from lightweight vision models to generative large models. QAIRT primarily
comprises two execution frameworks: SNPE and QNN. SNPE (Snapdragon Neural
Processing Engine) uses DLC (Deep Learning Container) format with good
cross-backend execution capability, suitable for rapid integration and prototype
verification. QNN is closer to the hardware execution layer, achieving efficient
deployment through platform-specific Context Binary generation, making it the
preferred path for production environments. Its deployment characteristics are
as follows.

\begin{itemize}
\item Precision and operator support are influenced by the target Snapdragon SoC
      model. Different chips' NPU/HTP capabilities and operator sets differ.
\item Context Binary generation of QNN is closely tied to the toolchain and
      system environment. For instance, for ARM64/Windows platforms, model
      conversion requires x64 emulation to complete compilation. For ARM/Linux
      platforms, the current version (v2.45) still lacks complete toolchain
      support, typically requiring development and compilation on x64/Linux
      hosts before deployment to target devices.
\item Dynamic shapes are not supported on the QNN HTP backend. All input
      dimensions must be fixed before deployment.
\item SDK version and target device compatibility require careful management.
\end{itemize}

Notably, while QAIRT SDK and ONNX Runtime QNN EP share the QNN library, they are
two independent deployment paths. First, QAIRT SDK provides QNN as a standalone
inference framework (ONNX\arw{}QNN Converter\arw{}Context
Binary) without depending on ONNX Runtime. Second, ONNX Runtime QNN EP is an
ONNX Runtime execution provider that achieves hardware acceleration by calling
the QNN library, preserving ONNX Runtime's inference pipeline and EP
architecture. The difference lies in deployment pipeline orchestration and
runtime integration level, not underlying hardware capabilities.

\subsection{Related Work for AI Agent Deployment}
\label{sec:bg:rw}
This paper builds on the prior AIPC framework~\cite{aipc}, which demonstrated
the feasibility of LLM agent-driven QAIRT model deployment. Recently, the Huawei
Ascend open-source community's agent-skills repository~\cite{ascendskills} also
adopted a similar ``skill + verification'' architecture, which provides the
\texttt{atc-model-converter} skill module for Ascend NPU, covering the
structured PyTorch\arw{}ONNX\arw\texttt{.om} (Offline
Model) conversion workflow. The \texttt{.om} is Ascend NPU's native inference
format, compiled from ONNX models by the Ascend Tensor Compiler, analogous to
RKNN's \texttt{.rknn} or TensorRT's \texttt{.engine}. Their shared practice
validates the ``skill + verification loop'' as a general design pattern for LLM
agent-driven deployment.

In related work on agent-based hardware deployment, most research focuses on
\textbf{subsystem optimization}, particularly operator and kernel-level
performance improvements, including AccelOpt~\cite{accelopt},
KForge~\cite{kforge}, Xe-Forge~\cite{xeforge}, EdgeFM~\cite{edgefm}, and
AVO~\cite{avo}. Li et al.~\cite{lispatialnpu} proposed a two-stage agent skill
system for AMD XDNA 2 NPU, distilling human expertise into structured
documentation for autonomous deployment of multiple LLMs.

Notably, \textbf{none of the above related work handles the complete end-to-end
PyTorch\arw{}ONNX\arw{}target format model conversion with
a deployment flow (pipeline) retargetable to various hardware platforms}. The
aforementioned kernel optimization works focus on performance improvement of
existing operators or kernels. Huawei Ascend's agent-skills~\cite{ascendskills}
is currently one of the few agent-driven works involving end-to-end conversion,
but covers only a single runtime. AIPC's differentiation lies in: systematizing
the PyTorch\arw{}ONNX\arw{}multiple target formats
conversion workflow, and exploring a unified deployment methodology across
heterogeneous runtimes through operator compatibility handling, layout
adaptation, and staged verification loops.

AIPC methodology further addresses two problems not explicitly solved in the
above works. (1) \textbf{Unsupported operator handling} provides substitution,
decomposition, and CPU fallback strategies for unsupported operators across
runtimes. (2) \textbf{Tensor layout adaptation} automatically handles
NCHW\,$\leftrightarrow$\,NHWC layout conversion at the ONNX level through
ONNXWrapper.

\subsection{Cross-Runtime Deployment Challenges and Motivation}
\label{sec:bg:chal}
Despite the variety of toolchains and APIs associated with different hardware
runtimes discussed in Sections~\ref{sec:bg:ov} to~\ref{sec:bg:qairt}, these
works commonly face four major challenges.

\textbf{(1) Operator Compatibility.} A subset of ONNX operators is supported by
all these runtimes. Unsupported operators require equivalent substitution or
computational graph rewriting before deployment. Furthermore, the set of
unsupported operators differs across runtimes, and substitution strategies vary
by runtime. In this case, it requires developers to master replacement knowledge
on a per-platform basis.

\textbf{(2) Tensor Layout Mismatch.} Conflicts exist between PyTorch's default
NCHW layout and the NHWC layout preferred by some runtimes. Unaddressed layout
mismatches cause conversion failures or silent performance degradation.

\textbf{(3) Dynamic vs.\ Fixed Shape Conflicts.} Research-oriented models often
use dynamic input dimensions, while edge inference runtimes generally prefer
fixed-shape compilation paths. Converting dynamic shapes to fixed shapes
requires understanding model semantics and hardware constraints, and is one of
the steps most frequently requiring manual intervention in deployment.

\textbf{(4) Quantization Dependence on Calibration Data and Environmental
Complexity.} Quantization is not simply an instruction to execute. The
representativeness of calibration data, consistency of preprocessing logic, and
the complexity of environments with coexisting multiple toolchain versions
(e.g., Windows on ARM, x86 emulation) affect the stability of quantization
results.

These common challenges mean that regardless of the target runtime, the
deployment workflow involves substantial engineering work requiring domain
knowledge, being repetitive and error-prone. Existing toolchains each address
format conversion problems individually, but lack a deployment automation
framework that can uniformly handle the above challenges across runtimes. This
is the motivation for the LLM agent-driven methodology in this paper. This
methodology explicitly encapsulates each runtime's domain knowledge through
skill modules, ensuring conversion quality through staged verification loops.
Moreover, our proposed methodology enables these agents to automatically handle
cross-platform common deployment challenges within constrained workflows.

\section{Agentic Deployment Methodology}
\label{sec:method}
In this work, an \textbf{agent} is an autonomous execution entity driven by
Large Language Models with tool-calling capabilities. During the edge AI
deployment process, the agent consists of three core components.
(1) \textbf{LLM backend} provides code generation, error analysis, and task
planning capabilities. (2) \textbf{Tool-calling interface} allows the agent to
execute shell commands, read/write files, and invoke external toolchains.
(3) \textbf{Skill modules} encapsulate runtime-specific domain knowledge
(toolchain commands, operator compatibility tables, common failure patterns) in
a structured manner to inject into the agent's execution context. In subsequent
sections, ``agent,'' ``LLM agent,'' and ``AI agent'' all refer to the same
concept and are used interchangeably depending on context.


\subsection{The ``Skill + Verification Loop'' Design Pattern}
\label{sec:method:loop}
Section~\ref{sec:bg:overview} outlines the staged pipeline for edge AI model
deployment (Model Source\arw{}ONNX Export\arw{}Runtime
Conversion\arw{}Verification\arw{}Deployable Artifact).
Notably, the Runtime Conversion stage can be further divided into single-stage
and dual-stage modes across different runtimes, distinguished by whether
intermediate executable artifacts exist. \textbf{Single-stage conversion}
directly generates the final deployable format from ONNX. That is, RKNN
generates \texttt{.rknn} in one step via RKNN-Toolkit2, and TensorRT generates a
serialized Engine (Plan file) in one step via \texttt{trtexec}.

\textbf{Dual-stage conversion} first produces an intermediate format, then
further compiles it into device-specific executable artifacts. OpenVINO first
generates device-agnostic IR (\texttt{.xml} + \texttt{.bin}) via
\texttt{ovc}/\texttt{mo}, and its runtime calls the function
\texttt{compile\_model()} to compile the intermediate file into a
device-specific model (with optional model cache). QAIRT first converts ONNX to
QNN Model (\texttt{.cpp}/\texttt{.bin}) as an intermediate artifact, and then the
intermediate files are compiled into a Context Binary (\texttt{.bin}).
Additionally, ONNX Runtime QNN EP provides more flexibility. It can directly load
ONNX models for execution via QNN EP (runtime compilation), or optionally enable
QNN context binary to serialize compilation results into an ONNX model with
\texttt{EPContext} nodes and a \texttt{.bin} file to accelerate subsequent
loading.

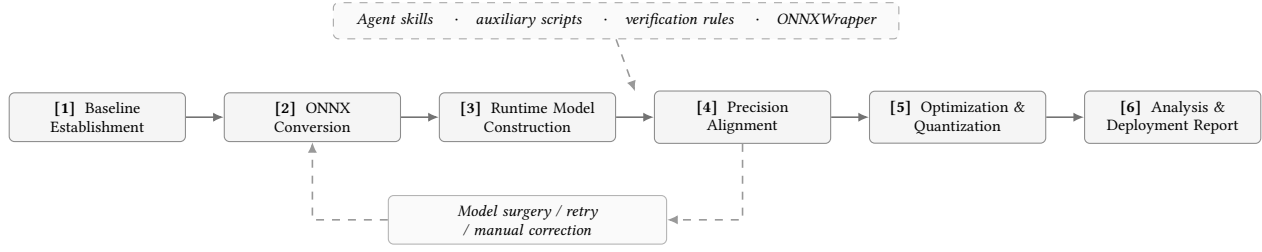
\begin{figure*}[t]
\centering
\begin{tikzpicture}[node distance=4mm and 5mm]
  \node[sstage]                    (s1) {\stepno{1} Baseline\\Establishment};
  \node[sstage, right=of s1]       (s2) {\stepno{2} ONNX\\Conversion};
  \node[sstage, right=of s2]       (s3) {\stepno{3} Runtime Model\\Construction};
  \node[sstage, right=of s3]       (s4) {\stepno{4} Precision\\Alignment};
  \node[sstage, right=of s4]       (s5) {\stepno{5} Optimization \&\\Quantization};
  \node[sstage, right=of s5]       (s6) {\stepno{6} Analysis \&\\Deployment Report};

  \draw[parrow] (s1) -- (s2);
  \draw[parrow] (s2) -- (s3);
  \draw[parrow] (s3) -- (s4);
  \draw[parrow] (s4) -- (s5);
  \draw[parrow] (s5) -- (s6);

  \node[band, above=7mm of s3.north east, anchor=south, text width=7.2cm]
       (sk) {Agent skills \quad$\cdot$\quad auxiliary scripts
             \quad$\cdot$\quad verification rules \quad$\cdot$\quad ONNXWrapper};
  \draw[farrow] (sk.south) -- ($(s3.north)!0.5!(s4.north)$);

  \node[rstage, below=7mm of s3, text width=3.4cm]
       (rp) {Model surgery / retry / manual correction};
  \draw[farrow] (s4.south) |- (rp.east);
  \draw[farrow] (rp.west) -| (s2.south);
\end{tikzpicture}
\caption{The ``skill + verification loop.'' The six-stage pipeline is wrapped by
injected domain knowledge (above) and a bounded repair path (below) that returns
a failed verification to an earlier stage rather than terminating the workflow.}
\label{fig:loop}
\end{figure*}

The AIPC framework~\cite{aipc} decomposes this common pipeline into six
verifiable stages, illustrated in Fig.~\ref{fig:loop}. (1) \textbf{Model
Preparation and Baseline Establishment} confirms the original model runs
correctly and establishes golden reference outputs. (2) \textbf{ONNX Conversion
and Modularized Inference Pipeline} exports the model to ONNX intermediate
representation as the common starting point for all runtimes. (3) \textbf{Runtime
Model Construction} uses target runtime conversion tools to generate deployable
formats. (4) \textbf{Inference Execution and Precision Alignment} verifies that
the converted model's output shapes, data types, and precision meet
expectations. (5) \textbf{Optimization and Quantization (Optional)} enters
optimization and quantization workflows only after the first four stages are
stable, avoiding unnecessary uncertainty. (6) \textbf{Performance Analysis and
Deployment Report} collects performance data and generates reusable deployment
documentation.

The core of AIPC lies not in the common pipeline itself, but in the ``skill +
verification loop'' mechanism wrapping the pipeline.

\begin{enumerate}
\item \textbf{Agent Skills} encapsulate runtime-specific knowledge: conversion
      commands, operator compatibility tables, quantization recipes, and common
      failure patterns with their fixes.
\item \textbf{Auxiliary Scripts} provide deterministic toolchain operations
      (model export, computational graph checking, inference benchmarking), which
      the agent invokes as tool calls.
\item \textbf{Verification Rules} define pass/fail criteria for each stage: ONNX
      model validity checks, conversion success/failure detection, output shape
      verification, and precision comparison with golden reference outputs.
\item \textbf{Failure Recoverability}. Model surgery, retry, and manual
      correction are integrated into the workflow, rather than treating failure
      as the end of the process. AIPC treats model surgery as a core capability
      of deployment automation rather than an exceptional case, with repair
      actions spanning three intervention levels. When ONNX export fails,
      modifying module implementations at the PyTorch source code level to
      eliminate operator or shape incompatibilities. When runtime conversion
      fails, patching nodes and attributes at the ONNX computational graph
      level. When runtime inference anomalies occur, performing further
      compatibility repairs based on actual errors, with fallback to earlier
      stages for reprocessing when necessary.
\item \textbf{ONNXWrapper Inference Replacement}. During the deployment
      verification stage, agents need to compare inference results against ONNX
      Runtime reference outputs for precision checking. \textbf{ONNXWrapper} is
      the inference replacement layer in AIPC, designed as a direct substitute
      for ONNX Runtime that additionally performs tensor layout adaptation.
      After replacing ONNX Runtime with ONNXWrapper in inference scripts,
      without modifying underlying code or regenerating target programs, the
      Wrapper automatically loads the corresponding hardware runtime model and
      produces output for comparison with ONNX Runtime results. This mechanism
      enables agents to rapidly complete multi-runtime precision verification
      without changing existing inference workflows. Its layout-adaptation role
      is detailed in Section~\ref{sec:pipe:wrapper}.
\end{enumerate}

\subsection{Agent Skills}
\label{sec:method:skills}
Table~\ref{tab:skills} lists the skill modules required by our proposed
methodology.

\begin{table}[t]
\caption{Skill modules required by the proposed methodology.}
\label{tab:skills}
\small
\begin{tabular}{@{}>{\raggedright\arraybackslash}p{0.30\columnwidth}
                   >{\raggedright\arraybackslash}p{0.60\columnwidth}@{}}
\toprule
\textbf{Skill Component} & \textbf{Description}\\
\midrule
Toolchain Reference           & CLI commands, API signatures, configuration formats\\
Operator Compatibility Matrix & Supported/unsupported operators for each runtime version\\
Quantization Guide            & Calibration procedures, precision trade-offs\\
Common Failure Patterns       & Error-to-fix mappings for common conversion issues\\
Verification Scripts          & Shape checkers, inference executors, precision comparators\\
Platform Constraints          & Memory limits, shape constraints, layout requirements\\
\bottomrule
\end{tabular}
\end{table}

\subsection{Single-Model Single-Runtime Orchestration}
\label{sec:method:orch}
The deployment scenario in this study is that given a PyTorch model and a target
runtime (OpenVINO, RKNN, TensorRT, or ONNX Runtime QNN), the agent is
responsible for handling the complete conversion workflow from the original
model to executable inference on that runtime. The agent's orchestration logic
is as follows.

\begin{enumerate}
\item \textbf{Runtime Identification}. Based on the user-specified target
      runtime, the agent loads the corresponding skill modules, such as
      toolchain commands, operator compatibility table, common failure patterns.
\item \textbf{ONNX Export}. The agent exports the PyTorch model to ONNX
      intermediate representation as the common starting point for all runtimes.
\item \textbf{Runtime-Specific Conversion}. The agent invokes the corresponding
      conversion tool based on the target runtime (OpenVINO \texttt{ovc},
      RKNN-Toolkit2, TensorRT \texttt{trtexec}, ONNX Runtime QNN EP), applying
      runtime-specific configuration parameters.
\item \textbf{Verification and Repair Loop}. The agent performs shape checking
      and inference precision verification on conversion results. If failures
      occur, the agent consults the repair patterns in the skill module to make
      corrections and re-verify.
\item \textbf{Deployable Artifact Output}. The agent generates model files that
      can be directly loaded and executed on the target runtime.
\end{enumerate}

Each deployment is an independent task. The agent focuses only on a single
model-to-single-runtime pairing, without cross-runtime parallel processing or
shared repairs.

\section{Runtime-Specific Deployment Pipelines}
\label{sec:pipe}
This section primarily explains that porting involves two parts, runtime-specific
code and ONNXWrapper. To validate the AIPC methodology, this study ports to a
single runtime at a time rather than simultaneously targeting multiple targets.
In implementation, our basic approach is to first use AI assistance to adapt the
AIPC Skill applicable to QAIRT, then through manual verification and
experimental adjustment, until the required functional modules can complete
testing.

Additionally, the design philosophy differs across hardware runtime toolchains.
TensorRT and RKNN conversion workflows typically do not rely on a common
intermediate representation. They directly generate independent execution
results for their respective hardware runtimes. In contrast, OpenVINO and QAIRT
have greater potential to support multiple hardware platforms through shared
intermediate formats. Different toolchains and hardware backends also differ
significantly in quantization and tensor layout support. In this case,
adaptation design must simultaneously address precision, quantization strategy,
and data layout conversion characteristics.

Table~\ref{tab:cross} summarizes the settings and operations performed by each
target runtime in the edge AI model deployment flow of Fig.~\ref{fig:common}.

\begin{table*}[t]
\caption{Cross-runtime mapping of the common deployment stages.}
\label{tab:cross}
\footnotesize
\setlength{\tabcolsep}{4pt}
\begin{tabular}{@{}>{\raggedright\arraybackslash}p{0.105\textwidth}
                   >{\raggedright\arraybackslash}p{0.155\textwidth}
                   >{\raggedright\arraybackslash}p{0.135\textwidth}
                   >{\raggedright\arraybackslash}p{0.165\textwidth}
                   >{\raggedright\arraybackslash}p{0.200\textwidth}
                   >{\raggedright\arraybackslash}p{0.175\textwidth}@{}}
\toprule
\textbf{Common Stage} & \textbf{OpenVINO} & \textbf{RKNN} & \textbf{TensorRT} &
\textbf{QAIRT (QNN)} & \textbf{ONNX Runtime QNN}\\
\midrule
Model Source & PyTorch & PyTorch & PyTorch & PyTorch & PyTorch\\
\addlinespace[2pt]
ONNX Export & \texttt{torch.onnx.export()} & \texttt{torch.onnx.export()} &
\texttt{torch.onnx.export()} & \texttt{torch.onnx.export()} &
\texttt{torch.onnx.export()}\\
\addlinespace[2pt]
Runtime Conversion &
\texttt{ovc}\,/\,\texttt{mo} $\rightarrow$ \textbf{IR} (\texttt{.xml} + \texttt{.bin}), optional model cache &
RKNN-Toolkit2 $\rightarrow$ \textbf{\texttt{.rknn}} &
ONNX Parser $\rightarrow$ Network Definition $\rightarrow$ Builder $\rightarrow$ \textbf{Engine/Plan} (\texttt{.engine}) &
\textbf{QNN Path}: QNN Converter $\rightarrow$ QNN Model, optional AOT compilation $\rightarrow$ \textbf{Context Binary} (\texttt{.bin}); \textbf{SNPE Path}: SNPE Converter $\rightarrow$ \textbf{DLC} (\texttt{.dlc}) &
ONNX Runtime + QNN EP direct execution (runtime compilation), optional \textbf{QNN context binary}\\
\addlinespace[2pt]
Conversion Mode &
Device-agnostic (IR) + device-specific (model cache) &
Device-specific (\texttt{.rknn}) &
Device-specific (Engine/Plan) + timing cache &
\textbf{QNN}: device-specific (Context Binary) + device-agnostic (QNN Model / \texttt{.so}); \textbf{SNPE}: device-agnostic (DLC) &
Device-agnostic (\texttt{.onnx}) + device-specific (QNN context binary)\\
\addlinespace[2pt]
Optimization \& Quantization &
NNCF PTQ: INT8/INT16 post-training quantization &
\texttt{do\_quantization} parameter, single-stage quantization &
\texttt{trtexec -{}-int8} build-time quantization &
QNN Converter built-in quantization or AIMET two-stage &
\texttt{onnxruntime.quantization} converts to QDQ, then hands off to QNN EP\\
\addlinespace[2pt]
Verification &
Cosine similarity, element-wise tolerance & Same as left & Same as left &
Same as left & Same as left\\
\bottomrule
\end{tabular}
\end{table*}

The following sections detail each runtime's specific deployment pipeline, agent
skill specifics, and common repair actions.

It is important to note that the model quantization is performed directly or
indirectly. The direct quantization is performed by the tools that use the
\textbf{single-stage conversion} scheme as described in
Section~\ref{sec:method:loop}. In the direct model quantization scheme,
quantization parameters are directly embedded in the conversion tool, with the
output being the quantized model (RKNN via \texttt{do\_quantization} parameter,
TensorRT via \texttt{trtexec -{}-int8} flag, QAIRT via QNN Converter quantization
options). The \textbf{dual-stage conversion} scheme adopts the indirect model
quantization. In this scheme, the model conversion is performed first, and then
the converted model independently runs the quantization process (i.e., OpenVINO
uses NNCF to quantize IR, ONNX Runtime QNN uses
\texttt{onnxruntime.quantization} to convert to QDQ model). Moreover,
quantization steps across runtimes can be divided into two categories:
\textbf{post-conversion independent quantization} (first converting to device
format, then quantizing that format) and \textbf{inline conversion
quantization} (quantization parameters specified in the conversion tool, output
is the quantized model).

The output formats of each runtime can be categorized as ``device-agnostic'' and
``device-specific,'' with most runtimes supporting both. Device-agnostic formats
(e.g., OpenVINO IR, ONNX) indicate the format is not bound to a single hardware
platform at the format level. It does not guarantee complete cross-device
portability, where different devices have different operator support ranges, and
some operators may only execute on specific devices. Device-specific formats
(e.g., TensorRT Engine, RKNN \texttt{.rknn}, QAIRT Context Binary) are bound to
specific hardware architectures and require recompilation for target devices.

\subsection{OpenVINO Deployment Pipeline}
\label{sec:pipe:ov}

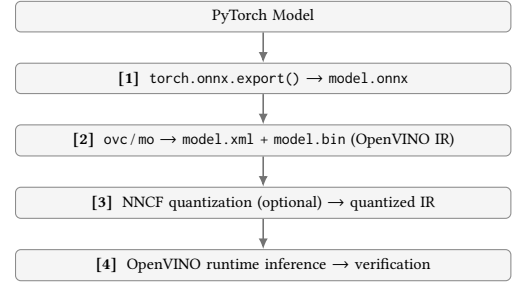
\begin{figure}[t]
\centering
\begin{tikzpicture}[node distance=4mm]
  \node[pstage]              (a) {PyTorch Model};
  \node[pstage, below=of a]  (b) {\stepno{1} \texttt{torch.onnx.export()} $\rightarrow$ \texttt{model.onnx}};
  \node[pstage, below=of b]  (c) {\stepno{2} \texttt{ovc}\,/\,\texttt{mo} $\rightarrow$ \texttt{model.xml} + \texttt{model.bin} (OpenVINO IR)};
  \node[pstage, below=of c]  (d) {\stepno{3} NNCF quantization (optional) $\rightarrow$ quantized IR};
  \node[pstage, below=of d]  (e) {\stepno{4} OpenVINO runtime inference $\rightarrow$ verification};
  \draw[parrow] (a)--(b); \draw[parrow] (b)--(c);
  \draw[parrow] (c)--(d); \draw[parrow] (d)--(e);
\end{tikzpicture}
\caption{OpenVINO deployment pipeline.}
\label{fig:ov}
\end{figure}

OpenVINO adopts the post-conversion independent quantization method. OpenVINO
has three implementation methods. (1) Traditional step-by-step approach. The
model is first converted to IR via \texttt{ovc}/\texttt{mo}, and then it runs
NNCF quantization script. (2) Combined API method. It sequentially calls
\texttt{convert\_model()} and \texttt{nncf.quantize()} in the same script, with
intermediate IR not saved to disk. (3) Integration packages are used for
quantization, such as \texttt{optimum-intel} for single-line conversion and
quantization. In contrast, RKNN, TensorRT, and QAIRT belong to the latter, with
quantization parameters directly written in conversion commands (e.g.,
\texttt{trtexec -{}-int8}, \texttt{rknn.config(do\_quantization=True)}), output
being the quantized model.

\textbf{Agent skill details.}
\begin{itemize}
\item \textbf{Export}. Agent uses OpenVINO-compatible settings (opset version,
      dynamic axis handling) to call \texttt{torch.onnx.export()}.
\item \textbf{Conversion}. Agent uses \texttt{ovc} (OpenVINO Converter) with
      target device specification (\texttt{CPU}, \texttt{GPU}, and \texttt{NPU}).
\item \textbf{Quantization}. Agent uses representative dataset for NNCF
      post-training quantization. Agent manages calibration data loading.
\item \textbf{Verification}. Agent uses cosine similarity and element-wise
      tolerance to compare OpenVINO inference output with PyTorch reference
      output.
\end{itemize}

\subsection{RKNN Deployment Pipeline}
\label{sec:pipe:rknn}

\begin{figure}[t]
\centering
\begin{tikzpicture}[node distance=4mm]
  \node[pstage]              (a) {PyTorch Model};
  \node[pstage, below=of a]  (b) {\stepno{1} \texttt{torch.onnx.export()} $\rightarrow$ \texttt{model.onnx} (simplified)};
  \node[pstage, below=of b]  (c) {\stepno{2} RKNN-Toolkit2 conversion (precision mode configurable) $\rightarrow$ \texttt{model.rknn}};
  \node[pstage, below=of c]  (d) {\stepno{3} RKNN runtime inference (on-device) $\rightarrow$ verification};
  \draw[parrow] (a)--(b); \draw[parrow] (b)--(c); \draw[parrow] (c)--(d);
\end{tikzpicture}
\caption{RKNN deployment pipeline.}
\label{fig:rknn}
\end{figure}
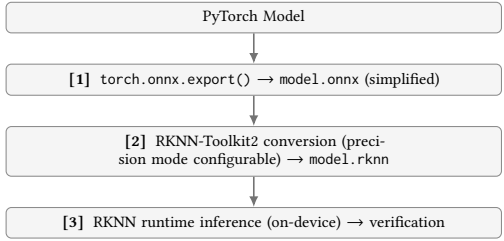

\textbf{Agent skill details.}
\begin{itemize}
\item \textbf{ONNX simplification}. Agent runs \texttt{onnx-simplifier} before
      RKNN conversion to clean the computational graph.
\item \textbf{Conversion and quantization}. Agent uses RKNN-Toolkit2 Python API,
      specifying platform configuration (\texttt{rk3588}) and
      \texttt{do\_quantization} parameter. Conversion and mixed quantization are
      completed in the same tool; agent manages per-layer quantization
      configuration and calibration datasets.
\item \textbf{Verification}. Agent performs on-device inference via RKNN
      runtime, and it compares output with reference values, accounting for
      quantization tolerance.
\end{itemize}

\subsection{TensorRT Deployment Pipeline}
\label{sec:pipe:trt}

\begin{figure}[t]
\centering
\begin{tikzpicture}[node distance=4mm]
  \node[pstage]              (a) {PyTorch Model};
  \node[pstage, below=of a]  (b) {\stepno{1} \texttt{torch.onnx.export()} $\rightarrow$ \texttt{model.onnx}};
  \node[pstage, below=of b]  (c) {\stepno{2} \texttt{trtexec} / TensorRT Builder (precision mode configurable) $\rightarrow$ \texttt{model.engine}};
  \node[pstage, below=of c]  (d) {\stepno{3} TensorRT runtime inference $\rightarrow$ verification};
  \draw[parrow] (a)--(b); \draw[parrow] (b)--(c); \draw[parrow] (c)--(d);
\end{tikzpicture}
\caption{TensorRT deployment pipeline.}
\label{fig:trt}
\end{figure}
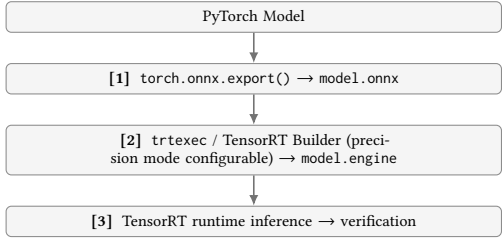

This study adopts the \textbf{ONNX\arw{}TensorRT} path (the first
deployment path described in Section~\ref{sec:bg:trt}), not the Torch-TensorRT
path.

\textbf{Agent skill details.}
\begin{itemize}
\item \textbf{Export}. ONNX exports with TensorRT-compatible opset and explicit
      batch dimension.
\item \textbf{Conversion and calibration}. Agent uses \texttt{trtexec} or Python
      TensorRT API, configuring builder settings (workspace size, FP16/INT8
      flags, dynamic shape optimization profiles). Quantization parameters are
      specified at build time; agent manages calibration cache generation and
      reuse.
\item \textbf{Verification}. Agent compares TensorRT output with PyTorch
      reference; handles numerical differences from reduced precision.
\end{itemize}

\subsection{ONNX Runtime QNN Deployment Pipeline}
\label{sec:pipe:ortqnn}
\begin{figure}[t]
\centering
\begin{tikzpicture}[node distance=4mm]
  \node[pstage]              (a) {PyTorch Model};
  \node[pstage, below=of a]  (b) {\stepno{1} \texttt{torch.onnx.export()} $\rightarrow$ \texttt{model.onnx}};
  \node[pstage, below=of b]  (c) {\stepno{2} QAIRT quantization (optional) $\rightarrow$ quantized \texttt{model.onnx}};
  \node[pstage, below=of c]  (d) {\stepno{3} ONNX Runtime + QNN EP execution $\rightarrow$ verification};
  \draw[parrow] (a)--(b); \draw[parrow] (b)--(c); \draw[parrow] (c)--(d);
\end{tikzpicture}
\caption{ONNX Runtime QNN deployment pipeline.}
\label{fig:ortqnn}
\end{figure}
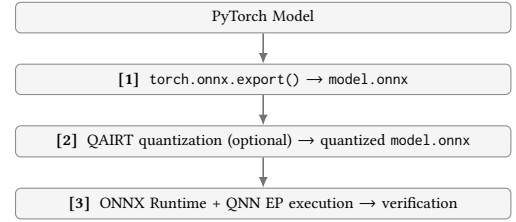

ONNX Runtime supports multiple hardware backends through the Execution Provider
architecture, including QNN EP (Qualcomm), OpenVINO EP (Intel), and TensorRT EP
(NVIDIA). Each EP handles device-specific compilation and caching. For example,
QNN EP provides \texttt{ep.context\_enable} to serialize compilation results as
QNN context binary, OpenVINO EP provides \texttt{cache\_dir} for caching
compiled models, and TensorRT EP provides engine cache mechanisms. This
architecture enables ONNX Runtime to use the unified ONNX format as a
device-agnostic intermediate representation while preserving device-specific
optimization capabilities through each EP. \textbf{This study focuses on the QNN
EP implementation}, with its deployment pipeline shown in Fig.~\ref{fig:ortqnn}.

ONNX Runtime QNN EP quantization belongs to the post-conversion independent
quantization method. That is, QNN EP itself does not handle quantization and
requires the input model to already be in QDQ format (ONNX model with
QuantizeLinear/DequantizeLinear nodes). Quantization must be completed
independently before QNN EP execution, using \texttt{onnxruntime.quantization}
tools to convert floating-point models to QDQ format before loading by QNN EP.

\textbf{Agent skill details.}
\begin{itemize}
\item \textbf{Export}. ONNX exports with QNN EP-compatible opset version; agent
      must confirm the opset support range corresponding to the target QAIRT SDK
      version.
\item \textbf{Quantization}. Optional use of QAIRT quantization toolchain for
      INT8/INT16 calibration on ONNX models. Agent manages calibration data
      loading and quantization parameter configuration.
\item \textbf{Execution}. Agent uses ONNX Runtime with QNN EP for inference;
      agent configures QNN backend path, sets HTP execution mode
      (\texttt{QNN\_HTP}, \texttt{QNN\_HTP\_FP16}), and device target (CPU/HTP).
\item \textbf{Verification}. Agent compares QNN EP inference output with PyTorch
      reference output, and handles numerical differences from quantization and
      NPU precision limitations.
\end{itemize}

\subsection{QAIRT Deployment Pipeline}
\label{sec:pipe:qairt}
QAIRT is the core target runtime of the original AIPC framework~\cite{aipc},
supporting both QNN and SNPE deployment paths.

\begin{figure}[t]
\centering
\begin{tikzpicture}[node distance=4mm]
  \node[pstage]              (a) {PyTorch Model};
  \node[pstage, below=of a]  (b) {\stepno{1} \texttt{torch.onnx.export()} $\rightarrow$ \texttt{model.onnx}};
  \node[pstage, below=of b]  (c) {\stepno{2} QNN Converter $\rightarrow$ QNN Model (\texttt{.cpp}/\texttt{.bin})};
  \node[pstage, below=of c]  (d) {\stepno{3} AOT compilation (optional) $\rightarrow$ Context Binary (\texttt{.bin})};
  \node[pstage, below=of d]  (e) {\stepno{4} QNN runtime inference $\rightarrow$ verification};
  \draw[parrow] (a)--(b); \draw[parrow] (b)--(c);
  \draw[parrow] (c)--(d); \draw[parrow] (d)--(e);
\end{tikzpicture}
\caption{QAIRT QNN deployment path, preferred for production.}
\label{fig:qnn}
\end{figure}
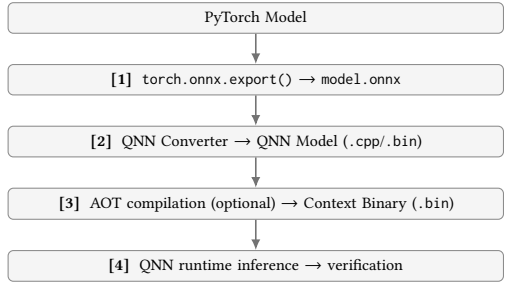

The QNN path, shown in Fig.~\ref{fig:qnn}, uses QNN Converter to convert ONNX
models to QNN Model (\texttt{.cpp}/\texttt{.bin}), which can be further compiled
via AOT compilation to generate Context Binary for accelerated loading. QNN
Model is loaded and executed by QNN backend libraries (\texttt{.so}/\texttt{.dll}),
supporting HTP (NPU), GPU, CPU, and other backends.

QAIRT supports two quantization modes. \textbf{Single-stage mode}: QNN Converter
has built-in quantization parameters, completing quantization during conversion.
\textbf{Two-stage mode}: It first uses AIMET (AI Model Efficiency Toolkit) for
ONNX model quantization (supporting PTQ and QAT), and then passes the quantized
model to QNN Converter for conversion. AIMET provides finer-grained quantization
control, suitable for scenarios with high precision requirements.

\begin{figure}[t]
\centering
\begin{tikzpicture}[node distance=4mm]
  \node[pstage]              (a) {PyTorch Model};
  \node[pstage, below=of a]  (b) {\stepno{1} \texttt{torch.onnx.export()} $\rightarrow$ \texttt{model.onnx}};
  \node[pstage, below=of b]  (c) {\stepno{2} SNPE Converter $\rightarrow$ DLC (\texttt{.dlc})};
  \node[pstage, below=of c]  (d) {\stepno{3} Model cache (optional) $\rightarrow$ device-specific cache file};
  \node[pstage, below=of d]  (e) {\stepno{4} SNPE runtime inference $\rightarrow$ verification};
  \draw[parrow] (a)--(b); \draw[parrow] (b)--(c);
  \draw[parrow] (c)--(d); \draw[parrow] (d)--(e);
\end{tikzpicture}
\caption{QAIRT SNPE deployment path, used for rapid prototype verification.}
\label{fig:snpe}
\end{figure}
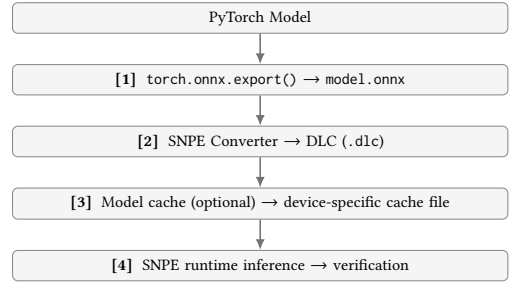

In the SNPE path of Fig.~\ref{fig:snpe}, the agent uses SNPE Converter to convert
ONNX models to DLC (Deep Learning Container) format. DLC is a device-agnostic
format that can execute across different backends (CPU, GPU, DSP) on the same
SoC, suitable for rapid verification and cross-backend testing. Additionally,
SNPE can optionally enable model cache to generate device-specific cache files
for accelerated loading, but its cache parameters differ from QNN's Context
Binary. QNN Context Binary requires actively specifying \texttt{-{}-soc\_version}
and other parameters for AOT compilation, while SNPE's model cache is
automatically managed by the runtime with fewer parameters.

\textbf{Agent skill details.}
\begin{itemize}
\item \textbf{Export}. ONNX export with QAIRT-compatible opset version.
\item \textbf{Conversion (QNN)}. Agent configures \texttt{qnn-onnx-converter}
      with \texttt{-{}-preserve\_io} to maintain I/O interface consistency.
\item \textbf{Conversion (SNPE)}. Agent configures SNPE Converter with
      quantization options.
\item \textbf{Quantization}. Agent uses QAIRT quantization toolchain for
      INT8/INT16 calibration; agent manages calibration data loading and
      quantization parameter configuration.
\item \textbf{Verification}. Agent compares QAIRT inference output with PyTorch
      reference output; handles numerical differences from quantization and NPU
      precision limitations.
\end{itemize}

\subsection{ONNXWrapper Inference Replacement and Layout Adaptation}
\label{sec:pipe:wrapper}
A recurring challenge in multi-runtime deployment is the mismatch between the
data layout assumed by model code (typically PyTorch's default NCHW) and the
layout required by certain hardware backends, along with optimization needs.
Manually inserting transpose operations for each model is tedious, error-prone,
and complicates the agent's repair logic.

For layout adaptation, ONNXWrapper, introduced in
Section~\ref{sec:method:loop}, analyzes the ONNX model layout and converts it as
needed by the target runtime (e.g., between NCHW and NHWC), while performing
shape consistency checks to verify conversion correctness. Because the Wrapper
sits at the inference-replacement boundary, this adaptation requires no change to
the original ONNX inference program logic.

\paragraph{QAIRT layout optimization}
In QAIRT's deployment implementation, ONNXWrapper's layout adaptation is treated
as part of the model optimization workflow. QNN models preserve layout options
during conversion (layout preservation), and ONNXWrapper compares layout
differences between the QNN backend and the original ONNX model, automatically
performing layout conversion during the optimization stage. Notably, such layout
conversion operations are more efficiently executed on CPU than on NPU, so
ONNXWrapper's conversion operations are scheduled on the CPU side to avoid
introducing additional computational overhead on the NPU.

\paragraph{RKNN layout adaptation}
The RK3588 NPU's compute units are optimized for NHWC tensor layout. When
standard NCHW ONNX models are directly passed to RKNN-Toolkit2, layout
mismatches occur and produce incorrect output results. ONNXWrapper automatically
pre-converts the computational graph to NHWC before conversion, thereby
eliminating layout-related failures. The ONNXWrapper method is integrated as a
pre-conversion step into the agent's RKNN deployment skill, automatically invoked
before RKNN-Toolkit2 conversion, reducing the agent's repair burden.

\section{Case Studies}
\label{sec:case}
This section evaluates the effectiveness of the agentic deployment method
through two sets of comparative experiments: (1) automated deployment using AIPC
agent skills and verification loops; (2) direct AI agent model conversion without
the AIPC framework, serving as a baseline control. Both experiments use the same
agent model (MIMO v2.5 / DeepSeek V4 Pro Preview) and target models, differing
only in whether the agent is injected with structured deployment domain knowledge
(operator compatibility tables, toolchain instructions, common failure patterns,
and staged verification rules). The experiments in this section only verify
deployment at NPU FP16 precision (non-TensorRT parts), without considering
performance, throughput, or other optimization factors, focusing on the
feasibility of the deployment workflow itself.

Field descriptions in the experimental tables are as follows.
\begin{itemize}
\item \textbf{End-to-end deployment time} includes agent thinking, code
      generation, toolchain execution, and verification. Due to differences in
      execution environments, LLM inference latency fluctuations, and varying
      numbers of AI steps, the same target's end-to-end deployment time may
      drift across experiments; this work did not perform performance
      optimization on the deployment environment.
\item \textbf{Repair actions} record specific repair measures taken by the agent
      when conversion fails.
\item \textbf{Human intervention count} measures the number of steps requiring
      manual correction, serving as the core metric for evaluating automation
      degree, including confirming or installing additional libraries, humans
      discovering and correcting workflow errors, AI halting and waiting for
      human intervention to resume, and humans proactively intervening to force
      error correction.
\item \textbf{Status} indicates deployment results (passed, partially passed,
      not completed).
\end{itemize}

\subsection{Experimental Setup}
\label{sec:case:setup}
This study selects two representative CV models as test targets:
\textbf{Real-ESRGAN}~\cite{realesrgan} and \textbf{YOLO-World}~\cite{yoloworld}.
Real-ESRGAN (Enhanced Super-Resolution Generative Adversarial Networks) is a
CNN-based image super-resolution model with regular structure and simple operator
types. It is primarily composed of convolutional and upsampling layers, making it
suitable as a baseline test case for deployment workflows. YOLO-World is an
open-vocabulary object detection model using a dual-encoder architecture (CLIP
text encoder and YOLO visual backbone). It requires multimodal inputs and more
complex operator combinations. These two models represent ``structurally regular'' and
``multi-encoder complex'' model categories, helping observe performance
differences of agent-based deployment across varying model complexities.

\textbf{Experimental prerequisites.} The following data should be prepared for each test model before the agent begins deployment to establish baselines. (1) Model
inference script should be present to ensure that the converted model runs correctly as in the original framework (PyTorch). (2) Model weights and (3) test inputs are required to ensure that the model can be executed as in the original framework. (4) Reference outputs are needed to verify that the model's output matches the expected results.
All models must be confirmed to run in the original environment before entering the deployment workflow to ensure baseline correctness.

The agentic deployment workflow in this case study primarily uses \textbf{MIMO
Code} (a CLI agent framework) for task orchestration and tool calls, driven by
\textbf{MIMO Auto} mode. The deployment uses the never-ask settings (no user interruption for queries), without additional skill installation or other configuration adjustments. Internally, it may dynamically switch between MIMO Pro and Flash
models. It should be noted that MIMO Code currently does not support Windows on ARM environments. For the QAIRT deployment target on Windows on ARM64 (Snapdragon X Plus), \textbf{Codex
CLI}~\cite{codexcli} with \textbf{DeepSeek V4 Pro Preview} is used as an
alternative agent-driven solution.

The execution environments for this study are as follows.
\begin{itemize}
\item \textbf{TensorRT}: GTX 1070 Ti, Ubuntu 24.04 Docker.
\item \textbf{OpenVINO}: Windows 11, Lunar Lake 228v, OpenVINO 2025.04.
\item \textbf{RKNN}: Ubuntu 24.04 WSL, Orange Pi 5 (RK3588S, Ubuntu 22.04),
      remote deployment using SSH.
\item \textbf{ONNX Runtime QNN}: Windows 11 ARM64, Snapdragon X Elite, QAIRT SDK
      2.45, ONNX Runtime 1.22, \texttt{onnxruntime\_qnn} 2.3.0,
      \texttt{olive-ai} 0.13.0.
\end{itemize}

\subsection{Experiment 1: AIPC Agentic Deployment}
\label{sec:case:exp1}
This experiment group uses the AIPC framework, including complete agent skill
modules (toolchain instructions, operator compatibility tables, common failure
patterns) and staged verification loops. The experimental workflow is driven by
the following three prompt instructions.

\begin{itemize}
\item \textbf{Inst. 1}: \texttt{Create AIPC project in this folder}. This instruction asks
      the agent to create an AIPC project in the specified folder.
\item \textbf{Inst. 2}: \texttt{[use provided env setting, fp16, <target>]. Auto
      fill the left fields and show the config for me to confirm.} This instruction provides the environment settings (FP16 precision, target runtime), with the agent
      auto-filling remaining configuration and displaying for confirmation.
\item \textbf{Inst. 3}: \texttt{Do all the work}. This instruction asks the agent to
      execute the complete deployment workflow and observe operation status.
\end{itemize}

This experiment group does not include QAIRT targets. As AIPC framework's
original validation runtime, QAIRT has been fully verified in prior
research~\cite{aipc} to complete end-to-end deployment with no (or near-zero)
human intervention. The goal of this experiment is to evaluate the feasibility of
extending the AIPC methodology to other heterogeneous runtimes (OpenVINO, RKNN,
TensorRT, ONNX Runtime QNN), thus focusing on these four new targets.

\subsubsection{Real-ESRGAN Super-Resolution}
Table~\ref{tab:e1esr} reports the results for Real-ESRGAN.

\begin{table*}[t]
\caption{Experiment 1 (with AIPC skill): Real-ESRGAN super-resolution.}
\label{tab:e1esr}
\footnotesize
\setlength{\tabcolsep}{5pt}
\begin{tabular}{@{}>{\raggedright\arraybackslash}p{0.16\textwidth}
                   >{\raggedright\arraybackslash}p{0.13\textwidth}
                   >{\raggedright\arraybackslash}p{0.28\textwidth}
                   >{\raggedright\arraybackslash}p{0.17\textwidth}
                   >{\raggedright\arraybackslash}p{0.14\textwidth}@{}}
\toprule
\textbf{Target} & \textbf{End-to-End Time} & \textbf{Repair Actions} &
\textbf{Human Intervention} & \textbf{Status}\\
\midrule
OpenVINO (NPU)          & 18 min & None & 0 & \yes{}Passed\\
RKNN (RK3588)           & 16 min & None & 0 & \yes{}Passed\\
TensorRT (FP16)         & 21 min & None & 0 & \yes{}Passed\\
ONNX Runtime QNN (HTP)  & 26 min & Olive optimization error, rename operator & 0 & \yes{}Passed\\
\bottomrule
\end{tabular}
\end{table*}

During ONNX Runtime QNN (HTP) execution, although we did not describe this error
handling in the skill, the agent handled the error autonomously in two steps.
First, Olive FP16 conversion created duplicate \texttt{Cast} output names, which
violates the single static assignment (SSA) property of the graph; the agent
renamed the duplicated node outputs. Second, the resulting nodes were no longer
topologically sorted after the SSA fix, and the agent applied a manual
topological sort.

\subsubsection{YOLO-World (Open-Vocabulary Detection)}
Table~\ref{tab:e1yolo} reports the results for YOLO-World.

\begin{table*}[t]
\caption{Experiment 1 (with AIPC skill): YOLO-World open-vocabulary detection.}
\label{tab:e1yolo}
\footnotesize
\setlength{\tabcolsep}{5pt}
\begin{tabular}{@{}>{\raggedright\arraybackslash}p{0.15\textwidth}
                   >{\raggedright\arraybackslash}p{0.11\textwidth}
                   >{\raggedright\arraybackslash}p{0.25\textwidth}
                   >{\raggedright\arraybackslash}p{0.24\textwidth}
                   >{\raggedright\arraybackslash}p{0.14\textwidth}@{}}
\toprule
\textbf{Target} & \textbf{End-to-End Time} & \textbf{Repair Actions} &
\textbf{Human Intervention} & \textbf{Status}\\
\midrule
OpenVINO (NPU) & 12 min &
Abandoned \texttt{onnxsim}, adopted direct conversion path &
1. Installed \texttt{onnxsim} dependency (later abandoned due to conversion failure) &
\yes{}Passed\\
\addlinespace[2pt]
RKNN (RK3588) & 25 min & None &
1. Paused after environment detection, waiting for user confirmation of remote deployment &
\yes{}Passed\\
\addlinespace[2pt]
TensorRT (FP16) & 17 min & None &
1. Agent did not strictly follow skill instructions, skipped ONNXWrapper step &
\yes{}Passed\\
\addlinespace[2pt]
ONNX Runtime QNN (HTP) & 21 min / 22 min &
Two tests: context binary generation fail (skip) / repaired \texttt{Einsum} operator & 0 &
\yes{}Partially passed, acceptable\\
\bottomrule
\end{tabular}
\end{table*}

During OpenVINO environment deployment, the agent attempted to use
\texttt{onnxsim} to simplify the ONNX computational graph, but the simplified
model failed at the conversion stage. The agent identified that the CLIP text
encoder's complex structure might exceed \texttt{onnxsim}'s processing
capability, so it abandoned the simplification step and used the unsimplified
ONNX model for direct conversion, ultimately completing deployment successfully.
This observation highlights the importance of pre-configuring the deployment
environment: if toolchain versions, dependency libraries, and target device
drivers are pre-installed, the agent's exploration time and failure risk during
environment preparation can be reduced.

During RKNN environment deployment, after completing environment detection, the
agent autonomously decided to upgrade the PyTorch version and completed inference
verification through the RKNN CPU simulator. However, the agent treated simulator
verification as a completion signal, pausing the workflow and asking the user
whether to proceed with the next step of remote device deployment. After manual
confirmation, the agent continued to complete RK3588 on-device deployment. The
final deployment passed verification, but analysis showed four operators fell
back to CPU execution, reflecting that YOLO-World's complex operator combinations
still have compatibility limitations on RKNN.

During TensorRT environment deployment, the agent exhibited two deviations from
skill instructions: (1) the agent once misjudged that the target hardware did not
support FP16 precision, autonomously switching to FP32 mode, then recovering FP16
settings after re-confirmation; (2) the agent skipped the ONNXWrapper layout
adaptation step specified in the AIPC skill module, directly executing TensorRT
conversion. These two deviation behaviors indicate that even within the
structured skill injection framework, agents may still deviate from the
predetermined workflow due to reasoning biases, emphasizing the necessity of
verification mechanisms and execution harnesses during deployment. Ultimately,
the agent successfully completed deployment in FP16 mode after correction.

In the ONNX Runtime QNN (HTP) environment, two tests were conducted, both facing
the same core challenge: the \texttt{Einsum} operator in the YOLO-World v2 model
is not supported by QNN EP, causing the computational graph to be split into six
independent \texttt{EPContext} nodes (rather than the expected single node). In
the first test, the agent attempted multiple times to generate QNN context binary
but all failed, ultimately choosing to skip this optimization step and execute
inference with the split computational graph; however, the AIPC skill module
actually already contained handling procedures for such situations, and the agent
failed to correctly trigger that repair path. In the second test, the agent
attempted equivalent replacement repair for the \texttt{Einsum} operator, replacing the \texttt{Einsum} node with an equivalent matrix multiplication
subgraph, ultimately completing deployment and passing verification.

\subsection{Experiment 2: Direct AI Conversion Without Structured Domain
Knowledge (Baseline Control)}
\label{sec:case:exp2}
This experiment group does not use the AIPC framework's skill modules and
verification loops; the AI agent directly attempts PyTorch to target runtime
conversion using only its general model knowledge. Compared to Experiment 1, the
agent is not injected with operator compatibility tables, toolchain
instructions, or common failure patterns. The purpose of this experiment is to
quantify the impact of structured domain knowledge injection on agent deployment
capability.

The test method is as follows. Given a PyTorch model and inference script, we use
the prompt ``please port the PyTorch \texttt{PROJECT} using \texttt{\{framework\}},''
with hints provided if environments are already set up.

\subsubsection{Real-ESRGAN Super-Resolution}
Table~\ref{tab:e2esr} reports the baseline results for Real-ESRGAN.

\begin{table*}[t]
\caption{Experiment 2 (without skill): Real-ESRGAN super-resolution.}
\label{tab:e2esr}
\footnotesize
\setlength{\tabcolsep}{5pt}
\begin{tabular}{@{}>{\raggedright\arraybackslash}p{0.15\textwidth}
                   >{\raggedright\arraybackslash}p{0.11\textwidth}
                   >{\raggedright\arraybackslash}p{0.25\textwidth}
                   >{\raggedright\arraybackslash}p{0.24\textwidth}
                   >{\raggedright\arraybackslash}p{0.14\textwidth}@{}}
\toprule
\textbf{Target} & \textbf{End-to-End Time} & \textbf{Repair Actions} &
\textbf{Human Intervention} & \textbf{Status}\\
\midrule
OpenVINO (NPU) & $\sim$19 min & None & 0 & \yes{}Passed\\
\addlinespace[2pt]
RKNN (RK3588) & $\sim$90 min &
\texttt{Reshape} operator repair, INT8 quantization adjustment, CPU simulator &
0, then prompted input conversion to NHWC; can complete but takes $\sim$2\,h &
\no{}Not completed\\
\addlinespace[2pt]
TensorRT (FP16) & $\sim$20 min & None & 0 & \yes{}Passed\\
\addlinespace[2pt]
ONNX Runtime QNN (HTP) & 20 min & Fixed shape & 0 &
\yes{}Passed (using older \texttt{onnxruntime-qnn} 1.x)\\
\addlinespace[2pt]
QAIRT QNN (FP16) & $\sim$60 min &
SDK structure analysis, SNPE workflow conversion & 0 &
\no{}Below expectations: converted from QNN to SNPE workflow\\
\bottomrule
\end{tabular}
\end{table*}

In this experiment, for TensorRT and OpenVINO NPU the agent could directly
understand the workflow and generate usable models. For ONNX Runtime QNN, the
agent used the older \texttt{onnxruntime-qnn} 1.x version, successfully running
Real-ESRGAN inference via QNN EP on HTP. The agent identified the QNN EP
limitation requiring fixed input dimensions for NPU targets during the process
and adjusted model configuration accordingly.

QAIRT QNN conversion required additional analysis of the SDK directory structure,
converting the model to the SNPE workflow and generating context binary, taking
approximately one hour. RKNN (RK3588) encountered output similarity mismatch with
PyTorch reference values during inference; the agent attempted modifying
operators and INT8 quantization configuration but could not complete, stopping
after one hour and reporting incomplete. Subsequent testing found that if
manually prompted that the hardware is NHWC format, RKNN conversion could pass.

Additionally, in the OpenVINO experiment, we discovered that on one occasion the
AI generated test images itself, ran through the workflow, then was asked to use
built-in test images, confirming functionality was correct.

Notably, in the ONNX Runtime QNN test, the agent autonomously selected the
deprecated \texttt{onnxruntime-qnn} 1.x version to complete conversion;
subsequent manual switching to the 2.x version caused failures. Considering the
agent successfully completed end-to-end inference under the 1.x version, this
test result was still marked as \textbf{usable (passed)}. This observation also
suggests that without skill explicitly specifying package versions, agents may
default to non-current stable versions, indicating a need to annotate recommended
version ranges in skill modules.

Additionally, without skill, we observed that the agent sometimes tends to
traverse the entire \texttt{onnxruntime-qnn} installation directory to locate
required libraries or configuration files, rather than directly accessing them
along predetermined paths. This extensive scanning behavior introduces a large
amount of unnecessary directory structure information, causing significant
additional input token consumption and correspondingly longer response times. In
contrast, skill, by providing explicit toolchain paths and API references, may
effectively suppress such inefficient exploration behavior, reducing the agent's
token cost and execution latency.

\subsubsection{YOLO-World (Open-Vocabulary Detection)}
Table~\ref{tab:e2yolo} reports the baseline results for YOLO-World.

\begin{table*}[t]
\caption{Experiment 2 (without skill): YOLO-World open-vocabulary detection.}
\label{tab:e2yolo}
\footnotesize
\setlength{\tabcolsep}{5pt}
\begin{tabular}{@{}>{\raggedright\arraybackslash}p{0.15\textwidth}
                   >{\raggedright\arraybackslash}p{0.11\textwidth}
                   >{\raggedright\arraybackslash}p{0.25\textwidth}
                   >{\raggedright\arraybackslash}p{0.24\textwidth}
                   >{\raggedright\arraybackslash}p{0.14\textwidth}@{}}
\toprule
\textbf{Target} & \textbf{End-to-End Time} & \textbf{Repair Actions} &
\textbf{Human Intervention} & \textbf{Status}\\
\midrule
OpenVINO (NPU) & 20 min & Dynamic shape to fixed shape & 0 & \yes{}Passed\\
\addlinespace[2pt]
RKNN (RK3588) & Not executed & Not executed & Not executed & Not executed\\
\addlinespace[2pt]
TensorRT (FP16) & 18 min & Dynamic shape to fixed shape & 0 & \yes{}Passed\\
\addlinespace[2pt]
ONNX Runtime QNN (HTP) & $\sim$60 min & --- &
2. Tested using non-context-binary mode, reported NPU inactivity &
\no{}Not passed\\
\addlinespace[2pt]
QAIRT QNN (FP16) & Not executed & Not executed & Not executed & Not executed\\
\bottomrule
\end{tabular}
\end{table*}

In this experiment, OpenVINO and TensorRT both passed; RKNN conversion previously
required manual prompting that the hardware is NHWC format, and QAIRT testing had
severe drift, so this section was not re-executed.

In the ONNX Runtime QNN experiment, we specified execution in a particular Python
virtual environment (venv) and explicitly instructed the agent to perform
inference in QNN Context Binary mode. The agent stopped after attempting to
execute Context Binary and failing; the user then issued instructions in
non-Context Binary mode, and the agent executed as instructed and reported
deployment success. However, actual verification revealed that the entire
inference process did not use NPU/HTP, and hardware monitoring tools confirmed
NPU was inactive throughout.

After feeding back the ``NPU inactivity'' observation to the agent, the agent
then identified the root cause and made corrections, ultimately completing
inference correctly executed on the HTP backend. This case illustrates that even
when the agent completes inference and reports success according to user
instructions, independent hardware activity verification is still needed to
confirm whether the target accelerator was actually activated, one cannot judge
deployment effectiveness solely based on the agent's report.

After the test failure, we attempted to reopen the conversation with the Claude
Sonnet 4.6 model (new conversation), letting the agent read artifacts and work
files left from the previous conversation as context, to re-attempt YOLO-World's
ONNX Runtime QNN Context Binary generation. This time, Context Binary generation
and inference verification were successfully completed in the new conversation.

\subsection{Analysis}
\label{sec:case:analysis}
Table~\ref{tab:summary} summarizes the observation metrics from both experiment
groups. It should be noted that this study's experimental sample size is limited
(4--8 deployment tests per group), and the following data are preliminary
observations only, not precise statistical analysis, lacking statistical
significance.

\begin{table}[t]
\caption{Summary of both experiment groups.}
\label{tab:summary}
\small
\setlength{\tabcolsep}{4pt}
\begin{tabular}{@{}>{\raggedright\arraybackslash}p{0.27\columnwidth}
                   >{\raggedright\arraybackslash}p{0.17\columnwidth}
                   >{\raggedright\arraybackslash}p{0.24\columnwidth}
                   >{\raggedright\arraybackslash}p{0.17\columnwidth}@{}}
\toprule
\textbf{Experiment Group} & \textbf{Success Rate} &
\textbf{Test Time (Successful)} & \textbf{Human Interv.}\\
\midrule
Exp.\ 1 (with AIPC skill) & 8/8 (100\%) & Avg.\ 19.5 min & 3\\
Exp.\ 2 (without skill)   & 3/5 (60\%)  & Avg.\ 19.8 min & 2\\
\bottomrule
\end{tabular}
\end{table}

``Deployment success rate'' is the number of passed verification deployments
divided by total attempts (excluding unexecuted targets); ``deployment test
time'' calculates only the average time of successful deployments, excluding
failed cases' times (as their duration is unrelated to deployment feasibility);
``human intervention count'' includes all human intervention events such as
confirming environments, installing dependencies, and suggesting correction
directions.

From this study's deployment results, for TensorRT and OpenVINO, AI agents
already demonstrate considerable understanding of common CV model deployment
tasks: they can handle ONNX conversion, dynamic shape profiles, and basic
deployment-time repairs, completing basic inference pipeline construction. RKNN's
layout requirements are more obscure, making it difficult for AI agents to fully
master all conversion conditions; in the cases, it was only after manual explicit
prompting to convert from NCHW to NHWC format that RKNN conversion could pass.
ONNX Runtime QNN benefits from the broad support of the standardized ONNX format,
with agents demonstrating considerable capability in basic conversion, but facing
challenges of limited QNN EP operator coverage (e.g., not supporting
\texttt{Einsum}) and Olive optimization tool integration, requiring agents to
handle operator equivalent substitution and context binary generation failures.

Synthesizing the above observations, the common problem across runtimes is that
the agent's failure modes are highly correlated with its command of deployment
knowledge. When operator support boundaries, layout requirements, and toolchain
behavior can be explicitly recorded and injected into the agent's context (as
provided by AIPC skill modules), the agent is better able to follow correct
repair paths; conversely, when this knowledge is scattered in documentation or
implicit in toolchain behavior, agents tend to deviate from workflows or adopt
inefficient trial strategies.

Synthesizing observations from both experiments, a more fundamental issue
emerges: \textbf{AI agents in engineering tasks exhibit a universal tendency
toward drift and hallucination}. Whether or not AIPC skill modules are used,
agents exhibit varying degrees of deviation behavior during deployment across
multiple runtimes, misjudging hardware capabilities (e.g., misjudging FP16
support in TensorRT environments), skipping critical steps (e.g., omitting
ONNXWrapper layout adaptation), and autonomously modifying environment
configurations (e.g., upgrading PyTorch version without authorization in RKNN
environments). Notably, \textbf{the longer the deployment workflow steps and the
more the operations deviate from conventional paths} (e.g., AIPC framework
requires layout adaptation via ONNXWrapper rather than directly calling hardware
runtime APIs), the more severe the agent's deviation becomes. The YOLO-World
deployment cases clearly demonstrate this: compared to the structurally regular
Real-ESRGAN, YOLO-World triggered agent deviation behavior across multiple
runtimes, even under AIPC skill injection constraints that could not be fully
avoided.

These observations indicate that AI agents in engineering practice should be
paired with appropriate verification mechanisms and staged checks, rather than
relying entirely on the agent's autonomous decision-making. The AIPC framework
demonstrated in this study serves as a preliminary porting verification tool; its
``skill + verification loop'' mechanism can constrain agent deviation behavior to
some extent, but still has shortcomings when facing complex models or
long-workflow tasks. In practical engineering deployment, more rigorous staged
verification, output comparison, and anomaly detection mechanisms are needed to
ensure each step of the agent executes on the expected trajectory.

\subsection{Observations and Limitations}
\label{sec:case:limits}
This section synthesizes experimental observations and methodological
limitations, explaining the current research coverage and unresolved issues.

\paragraph{Model complexity coverage boundary}
Current experiments only cover two CV models: Real-ESRGAN and YOLO-World.
Structurally regular convolutional networks (like Real-ESRGAN) have higher
conversion success rates across runtimes with fewer agent deviation behaviors;
models involving multi-encoder architectures (like YOLO-World) can still complete
deployment on most runtimes, but agent deviation behaviors are significantly more
frequent (e.g., skipping ONNXWrapper, misjudging hardware capabilities), and
operator fallback to CPU occurs on RKNN, reflecting that complex model structures
increase the difficulty of agents correctly following deployment workflows.
Transformer, LLM, and other more complex architectures' complete performance
across runtimes has not yet been systematically verified.

\paragraph{Cross-platform commonality and differences of operator repair not
fully clarified}
In handling unsupported operators, runtimes actually share multiple repair
mechanisms, operator substitution, CPU fallback, and plugin operator strategies
are all used in OpenVINO, RKNN, TensorRT, and QNN EP, with differences primarily
manifesting in trigger conditions, substitution strategies, and default behaviors
of each mechanism. Current experiments are case-by-case observations, lacking
systematic cross-platform comparison, and the applicable boundaries and best
practices of each repair mechanism across different runtimes cannot yet be
generalized.

\paragraph{Quantization and optimization not yet implemented}
This research currently focuses on deployment feasibility verification and has
not yet implemented quantization and optimization testing.

\paragraph{Agent model generalization boundary}
This study primarily uses MIMO v2.5 as the agent-driven model, only switching to
Codex CLI with DeepSeek V4 Pro Preview in the Windows on ARM environment (where
MIMO Code does not yet support the platform). Different agent models may have
significant differences in skill compliance, failure recovery capability, and
platform knowledge generalization, requiring larger-scale cross-model
verification.

\section{Conclusion}
\label{sec:concl}
This paper extends the AIPC agentic deployment methodology to multi-runtime
scenarios, including Intel OpenVINO, Rockchip RKNN, NVIDIA TensorRT, and ONNX
Runtime QNN. This method targets single-model-to-single-runtime pairings for automated
deployment. Preliminary results demonstrate the ``skill + verification loop''
design pattern's feasibility across different toolchains, where LLM agents
help with model conversion, operator repair, and precision verification under
structured knowledge injection. The ONNXWrapper layer removes tensor layout
adaptation from the agent's manual repair burden. We have conducted the controlled comparison between deployments performed with and without structured knowledge injection.
Our results show that the agent's failure modes can be attributed to knowledge availability rather than to model capability alone.

For the structurally regular vision models we tested, agents can successfully
complete deployment on most runtimes. On the other hand, general knowledge alone is insufficient for reliable automated handling when model structures become more complex (e.g., multi-encoder architectures). Examples of the complex situations include: target runtime operator coverage is limited (e.g., ONNX Runtime QNN not supporting \texttt{Einsum}), or runtimes have implicit tensor layout constraints (e.g., RKNN requiring NHWC format) in less common or complex scenarios. 



From a more fundamental perspective, hardware-specific inference runtimes are
inevitable on many edge platforms. Compared to generic PyTorch execution, using
native hardware runtimes (such as NPU and GPU dedicated inference engines)
typically brings significant performance improvements and power consumption
optimization. This is especially critical for resource-constrained edge devices.
In the past, deploying models to these underlying hardware required developers to
deeply understand each platform's toolchain, operator support range, quantization
schemes, and compilation configuration, a tedious and highly barrier-laden
process. As AI agent technology matures, this gap is being bridged: agents can
understand, orchestrate, and execute these complex hardware adaptation workflows,
making what was previously hardware deployment work achievable at lower cost and
in shorter time.

\bibliographystyle{ACM-Reference-Format}
\bibliography{aipc2}

\end{document}